# Effect of powder bed fusion process parameters on microstructural and mechanical properties of FeCrNi MEA: An atomistic study

Ishat Raihan Jamil[1], Ali Muhit Mustaquim[1], Mahmudul Islam[1], Md Shajedul Hoque Thakur[2] and Mohammad Nasim Hasan[1,*]

[1]Department of Mechanical Engineering, Bangladesh University of Engineering and Technology, Dhaka-1000, Bangladesh.

[2]Department of Materials Science and NanoEngineering, Rice University, Houston, TX, 77005, USA.

[*]Corresponding author *Email address:* nasim@me.buet.ac.bd

## Abstract

In our study, molecular dynamics (MD) simulations of laser powder bed fusion (LPBF) have been conducted on equimolar FeNiCr medium entropy alloy (MEA) powders. With the development of newer LPBF technologies capable of printing at the microscale, an even deeper understanding of the underlying atomistic effects of the process parameters on the microstructural and mechanical properties of the manufactured FeNiCr MEA products is required. In accordance with previous literature, the parameters of the LPBF process have been systematically varied, including layer resolution from 1 to 6, laser power from 100 µW to 220 µW, bed temperature from 300 K to 1200 K, and laser scan speed from 0.5 Å/ps to 0.0625 Å/ps. Consistent with prior macroscopic experimental findings, the atomistic results suggest that additive manufacturing using thinner layers imparts higher ultimate tensile strength (UTS) than fabricating with thicker layers. The latter, however, requires a shorter process time but induces keyhole defect formation if the laser-induced temperature is not sufficiently high enough. Increasing the temperature proves useful in mitigating this problem. Enhancement of UTS for the multi-rowed powders has been observed by raising the substrate temperature to 600 K or laser power to 160 µW during production. Beyond these critical limits, however, the UTS of the product diminishes due to the emergence of multiple vacancies. The results of our present study will help researchers to find a good balance between the production speed and strength of additive manufactured products at the nanoscale.

# 1. Introduction

The demand and utilization of 3D printing have been on the rise since printing machines became affordable and consumer-friendly. Additive Manufacturing (AM) is a 3D printing process in which an object is built up layer by layer through depositing materials according to the digital data provided [1–3]. Additive Manufacturing has revolutionized the manufacturing industry owing to its numerous benefits. AM reduces cost, time, and carbon footprint compared to the traditional methods of manufacturing. It minimizes raw material wastage in contrast to subtractive manufacturing processes and produces better quality parts for prototyping. It is possible to fabricate and customize complex geometries without the use of tools or molds. Using optimum parameters, the finished product can be manufactured with less porosity translating to better mechanical properties. Thus, it is widely used in aerospace, automotive, medical, and architectural industries [3–5].

Laser powder bed fusion (LPBF), one of the many 3D printing technologies, is being used to build metallic components with complex shapes [6, 7]. To perform LPBF, a layer of metal powder is placed over a substrate/bed. A high energy density laser is then used to selectively melt and fuse a specific region of the powder according to computer-aided design (CAD) data. The substrate is then lowered and another layer of powder is placed on top for the laser to scan again. Repeating this process enables the structure to be built following the design data. The final product has a near-net shape and up to 99.9% relative density [8]. Previous studies have reported various effects of different LPBF parameters on the structural integrity of numerous metals and alloys [9, 10]. It is reported that the LPBF process is associated with a high degree of porosity [11, 12]. These porosities can initiate micro-crack formation due to stress concentrations. Recently, studies have been performed on pore evolution to develop strategies to diminish porosity by varying process parameters. Aboulkhair et al. [13], for instance, improved the relative density of AlSi$_{10}$Mg alloy processed parts by using a trade-off between process parameters and scan strategies. On top of that, Maskery et al. [14] perceived correlations between scanning strategies, porosity levels, pore size, and their positions using X-ray computed tomography and have provided quantitative analysis methods to refine LPBF process parameters.

Numerous studies have adopted experimental and numerical methods to understand the physical phenomena like melt pool behavior and surface morphology during macroscale LPBF processing [15, 16]. In the LPBF process, melting and solidification occur at very small lengths and time scales. Thus, the high cost, low resolution, and complexity of measurement techniques make the experimental study of the thermo-fluid-mechanical behavior challenging. Also, numerical methods have been previously used to investigate the formation of LPBF products along with their melt pool geometry, temperature distribution, microstructural evolution, etc. Though continuum methods such as Computational Fluid Dynamics (CFD), Finite Element Method (FEM), Discrete Element Method (DEM) are used to study AM processes as well, working with the microscale powder bed fusion is more effective to understand the underlying physics during the LPBF process. Microscale-selective laser melting (µ-SLM) is a newly developing LPBF process, in which the melting phenomenon is dominated by nanoparticle composition, which the continuum models fail to present. Kurian et al. [17] in their study demonstrated that such emerging µ-SLM process could also be analyzed using molecular dynamics (MD) simulation at the nanoscale. As such, in this study, MD simulation would provide an exceptional way to visualize and understand why varying certain parameters can influence the overall mechanical and microstructural properties of a material.

Medium-entropy alloys (MEAs) such as FeNiCr are a newly emerging type of multicomponent alloys that have been developed recently. They are useful for applications such as dental implants [18], superconductivity [19], inert electrodes [20], etc. MEAs are defined as alloys that have a mixing entropy between 1R and 1.5R, where R is the gas constant [21]. MEA consists of 3 to 4 principle elements with equiatomic or near-equiatomic composition. MEAs are selected for the LPBF processes because of their superior mechanical properties, high thermodynamic stability, and excellent microstructure [22–24]. Yan et al. studied a medium-entropy alloy $Zr_{50}Ti_{35}Nb_{15}$ with a disordered body-centered cubic (BCC) structure which showed outstanding mechanical performances and corrosion resistance [25]. Recent advancements in the cost-effective production of equimolar Fe-Ni-Cr MEA with high tensile strength and ductility, by Liang et al. [26], make it an attractive choice for structural applications.

To understand the evolution of microstructure due to the LPBF process, Chen et al. [27] studied the atomic-scale crystallization and formation of equimolar FeCrNi MEA prepared by the LPBF process. They concluded that higher energy density and lower scanning speed can improve surface morphology. In their analysis, they have also found that Cr atom segregation occurs during the LPBF process but were able to control their size by varying the process parameters. Li et al. studied the effects of the solution and double aging treatment on the microstructures of the IN718 superalloy. Under the same aging condition, it is seen that the hardness of the alloy increased and then decreased with the increase of the solution temperature [28]. An experimental study by Sufiiarov et al. [29] found dependency between layer thickness and the strength of a nickel-based superalloy produced through the LPBF process. They concluded that higher layer thickness in the LPBF process produces lower strength properties. Tang et al. [30], in their study, reported that decreasing scan speed has a similar effect to that of increasing laser power and that for a single layer of stainless steel 316L, faster scan speed worsens the wettability of the powders over the substrate and gives rise to the interlayer defects with irregular patterns. Experimental observations by Aboulkhair et al. [13] also support their claim. Li et al. [31] experimentally studied the layer-by-layer construction of varying compositions of Fe-Ni-Cr based samples but using a melt deposition process. With the help of X-ray diffraction, they were able to identify different phases of the material in the manufactured product. All these studies described microstructural evolutions through the LPBF process. However, to the best of the authors' knowledge, a few questions remain to be answered, such as, which of these parameters has a more significant effect on the strength of the manufactured product at the nanoscale, and what are some of the other drawbacks, such as fabrication period. Although the strength of an additively manufactured product at the macroscale may not be completely consistent with that produced at the nanoscale using the same alloy due to numerous external factors, answering such questions would still provide some reasonable understanding of the atomistic effects varying the process parameters would have on the manufactured alloy product.

Taking motivation from the aforementioned research works, we have investigated the LPBF of multilayer equimolar FeNiCr MEA powders using molecular dynamics (MD) simulations in the present study. Process parameters such as layer resolution, laser power, bed temperature, and laser scan speed are systematically varied to analyze the particular effects they have on the strength properties of the manufactured products at the nanoscale.

## 2. Methodology

### *2.1 Atomic Structure Details*

Fig. 1 illustrates a simulation model consisting of 150 spherical equimolar FeNiCr MEA powders over a Fe substrate. The orientations of the simulation cell are considered to be X [1 0 0], Y [0 1 0], and Z [0 0 1]. Periodic boundary conditions are set in X and Y directions, whereas Z-direction has been kept free. Every row contains 25 spherical powders of radius 2.49 nm. Each equimolar powder is composed of 5727 Fe, Ni, and Cr atoms that are randomly distributed. The substrate bed, having a dimension of 25 nm × 25 nm × 1.7 nm, consists of 98000 Fe atoms in the FCC lattice which are divided into three sections along the Z direction: first, second, and third bed layers. The first bed layer atoms (Z: 0 nm to 0.2 nm) are fixed in place, whereas the motion of the second bed layer atoms (Z: 0.2 nm to 0.6 nm), third bed layer atoms (Z: 0.6 nm to 1.7 nm), and powder atoms are integrated using the Velocity-Verlet algorithm. The temperature of the second bed layer atoms, also referred to as the thermostat bed layer, is set using a Langevin thermostat. These atoms are responsible for heating the simulation cell to the required bed temperature as well as dissipating heat during cooling. Similar simulation setups were also adopted in previous studies [27, 32–35].

### *2.2 Interatomic Potential*

An embedded atom method (EAM) potential has been utilized to characterize the interatomic interaction forces. The potential developed by Bonny et al. [36] is used for this investigation since it has been successfully employed in the previous studies focusing on balling effect [27], melting metal powder [37], phase transformation, thermal conductivity, and vacancy formations [38–48]. In the EAM method, the total energy of the embedded atoms of the system is approximated as:

$$E_{total} = \sum_i \left[ F_i(\rho_i) + \frac{1}{2} \sum_{j(\neq i)} \Phi(R_{ij}) \right] \quad (1)$$

For atoms, $i$ and $j$, isolated by a distance of $R_{ij}$, $\Phi(R_{ij})$ represent the short-range electrostatic piecewise potential between them. $F_i$ is the embedding energy of atom $i$ in the host electron density which is denoted by the following equation:

$$\rho_i = \sum_{j(\neq i)} \rho_j^a(R_{ij}) \tag{2}$$

where the electron density $\rho_j^a$ at atom $i$ is due to another atom isolated by a length of $R_{ij}$.

*2.3 Simulation Procedures*

The calculations of the simulation cell are performed with a time step of 1 fs. The cell is equilibrated using the microcanonical ensemble (NVE) for 400 ps after adding new layers of powder. A cylindrical region of radius 5 nm [17, 27, 35] is modeled to add energy constantly to the system while moving along the Y-direction. This dynamic heating region simulates the motion of a laser beam during the LPBF process, which is depicted in fig. 1. Kurian et al. [17] bridged the gap between simulating µ-SLM at the microscale in FEA, and nanoscale in MD by approximating the melt pool temperature derived from the continuum approach. However, other relating studies [32–35] modeled the laser directly onto the MD domain thus opening up opportunities for exploring the effects of variation of its process parameters. The peak laser power density of some 3D laser metal sintering [49] systems can be as high as $2.8 \times 10^{12}$ W/m². Thus, with regards to the atomic scale laser spot of the current simulation domain, a maximum of 222 µW laser powers might be applicable. This range also encompasses the laser powers reported in the previously referenced literature, such as 80-128 µW utilized by Zhang et al. [35]. Additionally, these studies have also incorporated laser scan speeds up to of 2.0 Å/ps [32]. As such, the base velocity of the beam in this study is modeled to be 0.5 Å/ps, and once the powder bed fusion process is complete, the laser is turned off, and the system is allowed to cool down to the bed temperature.

After simulating all 6 rows of powder and cooling the simulation cell back to 300 K, a test zone of dimension 10 nm × 10 nm with a height of 9.6 nm (Z: 6.4 nm to 16 nm) is cut away from the middle of the cell. This region is common to all the simulations performed and ignores the first layer as it remains partially unmelted at higher layer resolutions or lower temperatures. Since the section is cut off from the main cell and is snipped to remove atoms in the surrounding regions, the shrink-wrapped boundary conditions are imposed, and the tensile test is carried out following the methods adopted by Kurian et al. [17]. The section is first relaxed by performing energy minimization using the conjugate gradient method. The block is then equilibrated at 300 K and 0 bar pressure along the Z-axis for 60 ps with a time step of 0.5 fs. An NPT ensemble is used for this process that employs Nose-Hoover style non-Hamiltonian equations of motion to perform time integration. After equilibration, a stain-controlled uniaxial tensile test along the Z-direction is performed by displacing two thin layers of atoms at the edges in opposite directions at a constant engineering strain rate of $10^9 \, s^{-1}$ up to 30% strain. The process summary is depicted in fig. 2.

All MD simulations are performed using the large-scale atomic/molecular massively parallel simulator (LAMMPS) [50]. The open visualization tool Ovito [51] is used for post-processing to visualize, perform common neighbor analysis (CNA), and dislocation analysis (DXA) from the data generated during MD simulations.

## 3. Results and Discussions

### *3.1 Process Parameter Overview and Control Simulation*

Table 1 summarizes the parameters varied in order to investigate the interrelationship between strength, process time, microstructure evolution for the LPBF of multilayer equimolar FeNiCr MEA. The term layer resolution is used to indicate the number of powder rows stacked over each other for the LPBF process. A low resolution of 1 means a single row of powders, representing a layer, is being melted by the laser at a time. After cooling, another row/layer is added, and the process is repeated until the laser has melted six layers. High resolution 6 means all six rows of powder stacked over each other are being acted upon by the layer at the same time. The summation of laser movement periods and the associated cooling intervals for each layer is considered to be the process time.

Following Kurian's method [17], a control simulation has been performed to determine the stress-strain relationship of a non-heat-treated equimolar FeNiCr MEA block at 300 K, having the same dimensions as described earlier. The equilibrated density of the material is determined to be about 8.16 g/cm$^3$. Fig. 3 shows the results of the uniaxial tensile test performed on the block. Stress appears to have a non-linear relationship with stain, being stiffer initially. Since non-periodic boundary conditions are used, the block oscillates when strained along the Z-axis, translating to an uneven stress-strain curve. Unstrianing, the block shows that the material is within the elastic limit up to the peak ultimate tensile strength (UTS) of 8.28 GPa at about 15.6% strain, after which necking occurs. This establishes a standard for the nanoscale multilayer LPBF simulation results to be compared to in the following sections.

**Table 1**. Details of variations of process parameters for each simulation.

| Case | Process Parameters | | | |
|:---:|:---:|:---:|:---:|:---:|
| | Layer Resolution | Laser Power (µW) | Bed Temperature (K) | Scan Speed (Å/ps) |
| A | 1 | 100 | 300 | 0.5 |
| B | 3 | 100 | 300 | 0.5 |
| C | 6 | 100 | 300 | 0.5 |
| D | 6 | 130 | 300 | 0.5 |
| E | 6 | 160 | 300 | 0.5 |
| F | 6 | 220 | 300 | 0.5 |
| G | 6 | 100 | 600 | 0.5 |
| H | 6 | 100 | 900 | 0.5 |
| I | 6 | 100 | 1200 | 0.5 |
| J | 6 | 100 | 300 | 0.25 |
| K | 6 | 100 | 300 | 0.125 |
| L | 6 | 100 | 300 | 0.0625 |

*3.2 Effect of Layer Resolution*

Experiments performed, at the macroscale, by Sufiiarov et al. [29] suggest a correlation between layer thickness and the strength of Inconel 718 produced through the LPBF process. Inconel 718 is primarily constituted of Fe, Ni, and Cr atoms, with trace elements such as Nb, Mo, and Ti, etc. However, Wang et al. [52] demonstrated that simulating it as a FeNiCr alloy with a greater nickel content produces good agreement with experimental observations. Based on their findings, three layer resolutions: 1, 3, and 6 are studied to gain atomistic insight into the effects of layer height on the LPBF of FeNiCr MEA. The bed temperature is set to 300 K and the laser is moved along the Y-direction at a speed of 0.5 Å/ps with a power of 100 µW. Fig. 4 depicts the layer addition mechanism for each resolution, while the end structures of the additive manufacturing process are shown in fig. 5. The post-cooling structures reveal that for resolution 1 the laser is able to effectively melt only the powders in its track without significantly affecting the surrounding ones. This signifies that smaller layer heights have good selective melting control. In the case of the other resolutions, the laser is not able to melt the bottommost row of powders completely.

Fe, Ni, and Cr form stable face-centered cubic (FCC) crystal lattices. The percentage of atoms in FCC structure within the isolated blocks is represented by the formula:

$$\%FCC = \frac{nFCC}{N_{atoms}} \times 100\% \tag{3}$$

where $nFCC$ = number of atoms in FCC lattice within the block and $N_{atoms}$ = total number of atoms in block. During the heating process, the powders lose their shapes resulting in an overall decrease in the final product's height compared to the starting material. The percentage difference between the equilibrated height of six unmelted rows of powders stacked over each other and the average height of solidified melt is referred to as shrinkage. It is calculated using the following formula:

$$\% Shrinkage = \frac{H - h}{H} \times 100\% \tag{4}$$

where $H$ = equilibrated height of 6 rows of unmelted powder over the bed and $h$ = average height of solidified melt above the same bed. It is of utmost importance for a designer to consider the material's shrinkage to ensure the dimensional accuracy of the final product. During the cooling stage, amorphously scattered atoms try to rearrange themselves into a stable FCC lattice. However, due to the interactions between different cooling fronts, dislocations form. The dislocation density of the blocks is defined as dislocation length per unit volume and is expressed by the formula:

$$Dislocation\ Density = \frac{L_d}{V} \tag{5}$$

where $L_d$ = total dislocation length within the block and $V$ = volume of the block. The partial dislocations develop stacking faults which results in the formation of hexagonal closed packed (HCP) structures, and the percentage of atoms in the HCP lattice within the isolated blocks is represented by the formulas:

$$\%HCP = \frac{nHCP}{N_{atoms}} \times 100\% \tag{6}$$

where $nHCP$ = number of atoms in HCP lattice within the block. The results of the uniaxial tensile test performed on the cut-off blocks as well as variation of the aforementioned properties of the simulations and blocks for the cases of varying layer resolutions are shown in fig. 6. It can be seen that the nanoscale simulation results suggest lower layer resolutions generate stronger material properties during the additive manufacturing process of FeNiCr MEA, similar to what Sufiiarov et al. [29] found experimentally for Inconel 718 at the macroscale. Thus, Wang's [52] determination of Inconel 718's representability as a FeNiCr alloy for molecular dynamic simulations, validates the findings of this nanoscopic study.

Just like the control blocks, the test zone blocks also exhibit nonlinear stress-strain relationships to some extent. Resolution 1 yields a comparable UTS of about 7.17 GPa to that of the control simulation but achieves peak strength only at about 9.8% strain. Resolution 3 follows closely behind, being about 10.2% weaker than the former. A combination of different sized isolated imperfections in the solidified resolution 3 simulation cell contributes to the lower %FCC, material density, and the greater dislocation density of the block. Resolution 6, on the other hand, manifests much lower UTS and density due to larger voids within the simulation block. It is visible from fig. 5(a), that increasing layer resolution tends to produce larger voids resulting in the decrease of both %FCC and material densities in fig. 6(c). As expected, a decrease in %FCC means an increase in %HCP since most of the atoms settle in one of these two lattice structures. It appears that the stacking faults containing atoms in the HCP lattice are more prominent near the voids in fig. 5(a). It suggests that the dislocation originates from these voids during the cooling process since the dislocation density shows an increasing trend similar to %HCP in fig. 6(c). It is perceptible from fig. 6(b) that the shrinkage decreases linearly with the increase in layer height, while the process time decreases at a diminishing rate. It is noteworthy that resolution 1 is about 2.3 times slower than resolution 3 and about 2.8 times slower than resolution 6.

While resolution 1 might seem more compelling due to better control over selective melting and higher UTS than resolution 3, the greater shrinkage and process time associated with it might suggest otherwise. On the contrary, resolution 6 might be the weakest but it has the least shrinkage and process time. All in all, resolution 3 achieves a good balance between the UTS and the process time. Although it is only about 18% slower than resolution 6 in this study, the actual process time gap between them will be much larger if the system is scaled up. As such, variations in laser power, scan speed, and bed temperature are investigated in the following sections in an attempt to mitigate the problems linked with the higher layer resolution, while considering simulation C as the baseline. Alleviating the imperfections of melting multi rowed powder structure would enable the use of thicker layers for the LPBF process, thereby lowering the production periods without compromising the UTS of the additively manufactured product.

*3.3 Effect of Laser Power*

Resolution 6, as compared to others, has a much lower material density and a greater dislocation density due to the development of large voids as discussed in the earlier section. For better understandings the defect evolution mechanisms, simulations are performed for 100 µW, 130 µW, 160 µW, and 220 µW laser powers, in accordance with some current machine settings [49] and prior literature [32–35] as previously mentioned, while keeping the layer resolution, laser scan speed, and bed temperature constant at 6, 0.5 Å/ps, and 300 K, respectively. Fig. 7 reveals the formation of voids with time for different laser powers in resolution 6 simulation cell. Keyhole defects are prominent at lower laser power, while numerous vacancies are observable for higher laser powers. Laser-induced keyhole defects, as visible for 100 µW, were also reported in previous studies involving LPBF and the laser welding process [53–56]. During the laser melting period, the melting front originates from the top and progressively moves downward and outward as the dynamic heating region passes rightward in the positive Y direction. The progression speeds of the fronts are noticeably higher at elevated temperatures induced by higher laser powers. As the laser moves forward, a void initiate from inter-powder spacing that gradually gets larger as the atoms of the melted powder coalescence to form the melt pool surrounding it. With time the temperature of the liquid melt continues to rise, shrinking the void in the process. The thermal energy of the melt is transferred downwards to be dissipated by the thermostat bed layer. The cooling front appears later in the heating stages, slowly moving upwards while settling the atoms in the melt pool into crystal lattices. The velocities of these fronts determine the cooling rate. A comparably low temperature, 1274 K, of the melt pool at the end of the heating stage for 100 µW laser means that the solidifying front moves fast enough to prevent the complete compression of the void, resulting in keyhole defect. However, if the melt pool temperature is too high, 1852 K as for the case of 220 µW laser beam, the cooling rate is too aggressive, leading to the formation of numerous vacancies in the FCC matrix. An optimum condition exists for 160 µW laser power in which the melt temperature and cooling rate are just sufficient to induce a minimum number of defects in the solidified alloy.

Fig. 8 shows the results of uniaxial tensile tests performed on the cut-off blocks as well as the properties of the same sections and simulations for increasing laser power. 130 µW laser-treated block has moderately better strength, material density, and half the dislocation density than that of 100 µW laser-treated block, owing to smaller void formations. 160 µW laser-treated block has the greatest UTS of 7.46 GPa at about 11% strain which is comparable to that of the control block. The increase in the UTS of 160 µW laser-treated block over the 100 µW laser-treated block is due to the presence of fewer vacancies as visible in fig. 7, an increase in %FCC and consequently a decrease in %HCP. The combination of a moderately high melt temperature of 1627 K coupled with a reasonably fast average cooling rate of 0.737 K/ps is enough to lower the viscosity sufficiently to allow the atoms to move more freely and arrange themselves in a stable FCC lattice within the test region during cooling. This results in a steady rise in %FCC from 100 µW to 160 µW after which it dips slightly, as visible in fig. 8(c). Material density follows a similar trend to that of %FCC but plateaus after 160 µW. A disproportionate distribution of Cr atoms could result in the density remaining unchanged from 160 µW to 220 µW even though numerous vacancies are present in the latter block, a phenomenon which is to be discussed in the next section. Better melting of the lower powders due to high thermal energy input, as can be seen from fig. 7, improves the contact area between the melt and the bed leading to faster cooling rates for 160 µW and 220 µW. It is observable from fig. 8(b) that higher laser power simulations have shorter process times, however, percentage shrinkage remains fairly unaffected. The 220 µW laser-treated block is much weaker than that of the 160 µW simulation, indicating the existence of a limiting laser energy density beyond which strength declines. Very high melt temperature generated by 220 µW laser power leads to a much faster average cooling rate of 0.913 K/ps that eventually lowers the %FCC of the block. The resulting increase in the presence of vacancies, as perceptible from fig. 7, weakens the overall strength of the block. It is noteworthy from fig. 8(c) that strength seems not to depend on the dislocation density as it appears to be higher for the best UTS scenario and lower for the other laser powers.

*3.4 Effect of Bed Temperature*

Increasing the substrate temperature might aid in melting and improving the microstructure at the lower end. To investigate this theory, the layer resolution, laser power, and speed are kept constant at 6, 100 µW, and 0.5 Å/ps, respectively, while the substrate temperature is varied from 300 K to 1200 K. Following the LPBF process, these systems are cooled in two phases. The melt is first allowed to cool to the bed temperature assigned for each case and equilibrate. Afterward, the thermostat bed layer atoms are set to 300K and the whole simulation cell is allowed to cool down further.

Results of the stain-controlled uniaxial tensile test of the isolated blocks and their properties for rising bed temperature are presented in fig. 9. The stress-strain graph suggests that increasing the bed temperature strengthens the laser-treated blocks, but up to a certain limiting temperature. 600 K bed temperature results in UTS of 7.84 GPa, at about 12.2% strain, which is almost 2 times stronger than that produced by 300 K. Its strength is very much similar to that of the control block, and the improvements are due to the alleviation of larger defects, as can be seen from fig. 10. Further temperature rise does not improve the strength and comes at the expense of greater Cr atoms segregation. Zeng et al. [57] observed that the large self-diffusibility of Cr prompts the Cr atom to migrate towards the surface as temperature rises, with smaller clusters forming within the nanoparticle. Experimental study of surface tension of molten Ni-(Cr, Co, W) by Xiao et al. [58] also reported similar occurrences. It is evident from fig. 10 and from the simulation findings of Chen et al. [26], that a similar phenomenon also occurs during the LPBF process of FeNiCr MEA. Higher bed temperature promotes larger Cr cluster formation, thus leading to disproportionate distribution of elements. Variation in material densities are generally indicative of the presence and sizes of voids, however, it is also affected by the uneven distribution of Cr atoms since they have lower atomic masses as compared to Fe and Ni atoms. Greater atomic diffusions associated with Cr atoms segregation at higher temperatures lead to increased dislocation densities. Contrary to increasing laser power, however, dislocation density appears to be minimum at the best UTS case conditions.

From fig. 9(c) it can also be seen that %FCC rises and peaks at 600K with a subsequent decrease at even higher bed temperatures. Similar to the previous sections, a higher %FCC is associated with a greater strength of the block. Percentage shrinkage increases marginally but proportionally with temperature, owing to proper melting of the bottommost powders. It takes more time for 600 K simulation to dissipate the heat to the thermostat bed layer atoms as compared to 900 K simulation even though its temperature is lower. Again, better bed surface wetting at higher temperatures, as seen from fig. 10, is critical for aggravating the average cooling rate from 0.499 K/ps for 300 K to 0.665 K/ps for 900 K, leading to decreased process time and increased dislocation density. For 1200 K, however, the average cooling rate is about 0.574 K/ps. The temperature of the melt is too high even for full surface wetting to cool it fast enough. Further increase in bed temperature above 900 K is not recommended since the material starts to melt just over 1000 K. Thus, partial melting of the powder, even without the laser, would render the selective melting control useless.

*3.5 Effect of Laser Scan Speed*

Increasing the scanning period by reducing the laser speed can be yet another way of improving the microstructure and strength of the final product constructed using the LPBF process. For this investigation, the scanning speed is decreased from a fast 0.5 Å/ps speed to a slower velocity of 0.0625 Å/ps, while keeping the layer resolution, laser power, and bed temperature constant at 6, 100 µW, and 300 K respectively. Results of the stain-controlled uniaxial tensile test and various properties of the simulations and blocks for decreasing scan speed are presented in fig. 11. Slowing down the laser increases the process time significantly and marginally improves the %FCC, while shrinkage remains fairly unaffected. These MD results correspond to the finite element analysis (FEA) observation made by Tang et al. [30] that decreasing scan speed has a similar effect on the microstructure to that of increasing laser power for stainless steel 316L, an alloy having similarly high composition of Fe, Ni, and Cr, to the MEA in this study. However, the same cannot be said for the case of UTS. Uniaxial tensile test results reveal that decreasing the speed from 50 Å/ps to 0.25 Å/ps raises the strength by 92.7%. 0.0625 Å/ps laser speed produces the greatest UTS of 7.50 GPa at about 11.5% strain, with 0.125 Å/ps laser speed produced block close behind being only about 3.6% weaker. Both blocks exhibit strength similar to that of the control block. But unlike increasing laser power or bed temperature, the UTS gains by the speed reduction for FeNiCr MEA seem not to diminish beyond a certain velocity, within the simulated range—the only drawback being the long process times required. Since the UTS attained at 0.0625 Å/ps is already comparable to that of an untreated block of FeNiCr MEA, further simulations at lower velocities are not performed.

Similar to elevated bed temperatures, prolonged heating periods also encourage more atomic diffusion leading to larger Cr cluster formation, as seen from the element distribution in fig. 12. For reduction in speed, dislocation density does not seem to follow any distinguishable trend, as can be seen from fig. 11(c), and does not appear to have an effect on the %HCP of the blocks. However, DXA in fig. 12 shows that the HCP stacking faults generate along several planes, originating from and ending at the dislocation lines. With fewer dislocation lines available for termination within the blocks, 0.25 Å/ps and 0.0625 Å/ps speeds generate a large percentage of HCP stacking faults in an FCC crystal lattice as compared to the block produced using 0.125 Å/ps laser speed. Such dependency could be attributed to the nature of various types of dislocations, such as Shockley, Stair-rod, and Hirth dislocations as shown by the DXA in fig. 12. Shockley partials are a pair of dislocations that cause the formation of stacking faults. Two pairs of mobile Shockley partials on two slip systems interacting at the intersecting junction of the planes result in the generation of sessile stair-rod dislocation. Similarly, when two glissile perfect dislocations on two interacting slip planes meet at their junction Hirth lock is formed if the summation of their burgers vector is of type <100> [59]. The reason why the ultimate tensile strengths obtained from the MD simulation do not agree fully with the similarity of effects observed by Tang et al. [30] for the variation of scan speed and laser power in the FEA simulation results could be due to the effects of the various dislocations on the microstructural strengths, which the continuum approach fails to capture. It is clear from the discussions that UTS is positively correlated to %FCC and subsequently inversely related to %HCP but doesn't seem to be dependent on the overall dislocation length. As such, further studies are required to investigate the influence of the various different types and lengths of dislocations on the UTS of products produced through the LPBF process. The highest %HCP is realized for case C due to the presence of large voids as discussed earlier.

*3.6 Comparing the Best Scenarios*
Fig. 13 summarizes, the percentage increase in both UTS and process time for the best strength scenarios of the nanoscale alloy product from each section as compared to the baseline simulation C product. The greatest improvement in atomistic UTS, 101%, is observed just by elevating the bed temperature from 300 K to 600 K while keeping other process parameters the same as case C. However, it comes at the cost of an 11% increase in process time. Raising the laser power to 160 µW from 100 µW in case C strengthens the product by 92% with the added benefits of a 19% reduction of process time. Layer resolution 1, as compared to resolution 6 and the same parameters as case C, shows the least enhancement of strength at about 84%. It also comes at the expense of a staggering 177% surge in process time. Slowing down the laser from 0.5 Å/ps to 0.0625 Å/ps exhibits the largest aggravation in process time by about 92%, with an impressive 203% amplification in UTS. It is noteworthy that these optimum parametric values are only relative to the scale of the simulation domain. As mentioned earlier, the macroscopic experimental results might not be fully consistent with those of the nanoscopic simulations due to the influence of several external factors. However, these MD findings might suggest some starting points and basis of explanations for further microscopic experimentations such as with µ-SLM. For instance, as mentioned above, the rise in bed temperature to 600 K offers the best UTS with a small increase in process time. This interpretation, however, does not account for the additional time and energy required for rising and maintaining the entire bed along with the alloy material to a higher temperature. Thus, in view of practical time-sensitive production processes, such excess requirements over the strength gains might make increasing bed temperature more undesirable. Hence, from the atomistic perspective, it might be more fruitful to further investigate and experiment with various laser configurations to better understand their underlying impacts on the alloy for additive manufacturing production, rather than elevating the bed temperature or using lower layer resolutions.

## 4. Conclusion

Our present molecular dynamic study explored the effects of process parameters such as layer resolution, laser power, bed temperature, and scan speed on the LPBF process of equimolar FeNiCr MEA. The conclusions of the study can be summarized as follows:

- Additive manufacturing, from the nanoscopic perspective, using a layer resolution of 1 provides a UTS of 7.17 GPa and reasonably good selective melting control, the downside being greater shrinkage and long process times required. Increasing the layer resolution mitigates these problems but induces the formation of large defects that significantly weakens the manufactured product.

- For production with a layer resolution of 6, elevating the bed temperature to 600 K achieves a UTS of 7.84 GPa which is very close to the strength of non-heat treated FeNiCr MEA. A comparable increase in strength is also observed by raising the laser power to 160 µW without altering the substrate temperature. Both cases require shorter process periods and would allow the use of thicker layers for additive manufacturing without compromising the ultimate tensile strength. Upsurge in either quantity beyond these critical values has the same effects as reducing them, which is weakening of the manufactured product.
- Slowing down the laser scanning speed also aids in strengthening the manufactured product but without the presence of any observable limiting conditions within the simulated range. It, however, suffers from extended production periods, similar to that of the lower layer resolutions. For layer resolution 6, the highest UTS obtained was 7.50 GPa at 0.0625 Å/ps which is very close to that of an untreated block of FeNiCr MEA, implying that any slower speed is not required.
- Higher bed temperatures and slower scanning speeds encourage greater Cr segregation resulting in the disproportional distribution of Cr, altering the material density at different locations in the final product.

This atomistic insight of mechanical and microstructural characteristics through multilayer LPBF process could aid in the production of FeNiCr MEA products with high UTS values at reasonable production speeds using upcoming additive manufacturing technologies such µ-SLM. Further simulation studies with different powder sizes and alloys, as well as microscopic experimentations could aid in process parameter optimization for the fast-growing and time-sensitive 3D printing technologies.

**Declaration of competing interest**

The authors declare that they have no known competing financial interests or personal relationships that could have appeared to influence the work reported in this paper.

**Acknowledgement**

The authors would like to acknowledge Multiscale Mechanical Modeling and Research Networks (MMMRN) for their technical assistance to carry out the research. The authors also acknowledge the high performance computing facilities provided by the Institute of Information and Communication Technology (IICT), BUET during this study.

**References**

1. Ngo TD, Kashani A, Imbalzano G, et al (2018) Additive manufacturing (3D printing): A review of materials, methods, applications and challenges. Compos Part B Eng 143:172–196. https://doi.org/10.1016/j.compositesb.2018.02.012
2. DebRoy T, Wei HL, Zuback JS, et al (2018) Additive manufacturing of metallic components – Process, structure and properties. Prog Mater Sci 92:112–224. https://doi.org/10.1016/j.pmatsci.2017.10.001
3. Frazier WE (2014) Metal additive manufacturing: A review. J Mater Eng Perform 23:1917–1928. https://doi.org/10.1007/s11665-014-0958-z
4. Ford S, Despeisse M (2016) Additive manufacturing and sustainability: an exploratory study of the advantages and challenges. J Clean Prod 137:1573–1587. https://doi.org/10.1016/j.jclepro.2016.04.150
5. Durakovic B (2018) Design for additive manufacturing: Benefits, trends and challenges. Period Eng Nat Sci 6:179–191. https://doi.org/10.21533/pen.v6i2.224
6. Santos EC, Shiomi M, Osakada K, Laoui T (2006) Rapid manufacturing of metal components by laser forming. Int J Mach Tools Manuf 46:1459–1468. https://doi.org/10.1016/j.ijmachtools.2005.09.005
7. Pauly S, Löber L, Petters R, et al (2013) Processing metallic glasses by selective laser melting. Mater Today 16:37–41. https://doi.org/10.1016/j.mattod.2013.01.018
8. Yap CY, Chua CK, Dong ZL, et al (2015) Review of selective laser melting: Materials and applications. Appl Phys Rev 2:. https://doi.org/10.1063/1.4935926
9. Kasperovich G, Haubrich J, Gussone J, Requena G (2016) Correlation between porosity and processing parameters in TiAl6V4 produced by selective laser melting. Mater Des 105:160–170. https://doi.org/10.1016/j.matdes.2016.05.070
10. Shipley H, McDonnell D, Culleton M, et al (2018) Optimisation of process parameters to address fundamental challenges during selective laser melting of Ti-6Al-4V: A review. Int J Mach Tools Manuf 128:1–20. https://doi.org/10.1016/j.ijmachtools.2018.01.003
11. Parimi LL, Ravi G, Clark D, Attallah MM (2014) Microstructural and texture development in direct laser fabricated IN718. Mater Charact 89:102–111. https://doi.org/10.1016/j.matchar.2013.12.012
12. Thijs L, Verhaeghe F, Craeghs T, et al (2010) A study of the microstructural evolution during selective laser melting of Ti-6Al-4V. Acta Mater 58:3303–3312. https://doi.org/10.1016/j.actamat.2010.02.004
13. Aboulkhair NT, Everitt NM, Ashcroft I, Tuck C (2014) Reducing porosity in AlSi10Mg parts processed by selective laser melting. Addit Manuf 1:77–86. https://doi.org/10.1016/j.addma.2014.08.001
14. Maskery I, Aboulkhair NT, Corfield MR, et al (2016) Quantification and characterisation of porosity in selectively laser melted Al-Si10-Mg using X-ray computed tomography. Mater Charact 111:193–204. https://doi.org/10.1016/j.matchar.2015.12.001

15. Yu G, Gu D, Dai D, et al (2016) On the role of processing parameters in thermal behavior, surface morphology and accuracy during laser 3D printing of aluminum alloy. J Phys Appl Phys 49:135501. https://doi.org/10.1088/0022-3727/49/13/135501
16. Wu YC, San CH, Chang CH, et al (2018) Numerical modeling of melt-pool behavior in selective laser melting with random powder distribution and experimental validation. J Mater Process Technol 254:72–78. https://doi.org/10.1016/j.jmatprotec.2017.11.032
17. Kurian S, Mirzaeifar R (2020) Selective laser melting of aluminum nano-powder particles, a molecular dynamics study. Addit Manuf 35:101272. https://doi.org/10.1016/j.addma.2020.101272
18. Wang S, Wu D, She H, et al (2020) Design of high-ductile medium entropy alloys for dental implants. Mater Sci Eng C 113:110959. https://doi.org/10.1016/j.msec.2020.110959
19. Lee YS, Cava RJ (2019) Superconductivity in high and medium entropy alloys based on MoReRu. Phys C Supercond Its Appl 566:1353520. https://doi.org/10.1016/j.physc.2019.1353520
20. Ndong GK, Xue J, Feng L, Zhu J (2015) Effect of anodic polarization on layer-growth of Fe-Ni-Cr anodes in cryolite-alumina melts. TMS Annu Meet 2015-March:83–90. https://doi.org/10.1002/9781119093381.ch11
21. Carroll R, Lee C, Tsai CW, et al (2015) Experiments and Model for Serration Statistics in Low-Entropy, Medium-Entropy, and High-Entropy Alloys. Sci Rep 5:1–12. https://doi.org/10.1038/srep16997
22. Gali A, George EP (2013) Tensile properties of high- and medium-entropy alloys. Intermetallics 39:74–78. https://doi.org/10.1016/j.intermet.2013.03.018
23. Yang M, Zhou L, Wang C, et al (2019) High impact toughness of CrCoNi medium-entropy alloy at liquid-helium temperature. Scr Mater 172:66–71. https://doi.org/10.1016/j.scriptamat.2019.07.010
24. Praveen S, Bae JW, Asghari-Rad P, et al (2018) Ultra-high tensile strength nanocrystalline CoCrNi equi-atomic medium entropy alloy processed by high-pressure torsion. Mater Sci Eng A 735:394–397. https://doi.org/10.1016/j.msea.2018.08.079
25. Yan X, Zhang Y (2020) A body-centered cubic Zr50Ti35Nb15 medium-entropy alloy with unique properties. Scr Mater 178:329–333. https://doi.org/10.1016/j.scriptamat.2019.11.059
26. Liang D, Zhao C, Zhu W, et al (2019) Overcoming the strength-ductility trade-off via the formation of nanoscale Cr-rich precipitates in an ultrafine-grained FCC CrFeNi medium entropy alloy matrix. Mater Sci Eng A 762:138107. https://doi.org/10.1016/j.msea.2019.138107
27. Chen H, Fang Q, Zhou K, et al (2020) Unraveling atomic-scale crystallization and microstructural evolution of a selective laser melted FeCrNi medium-entropy alloy. CrystEngComm 22:4136–4146. https://doi.org/10.1039/d0ce00358a
28. Li J, Zhao Z, Bai P, et al (2019) Microstructural evolution and mechanical properties of IN718 alloy fabricated by selective laser melting following different heat treatments. J Alloys Compd 772:861–870. https://doi.org/10.1016/j.jallcom.2018.09.200
29. Sufiiarov VS, Popovich AA, Borisov EV, et al (2017) The Effect of Layer Thickness at Selective Laser Melting. Procedia Eng 174:126–134. https://doi.org/10.1016/j.proeng.2017.01.179
30. Tang C, Tan JL, Wong CH (2018) A numerical investigation on the physical mechanisms of single track defects in selective laser melting. Int J Heat Mass Transf 126:957–968. https://doi.org/10.1016/j.ijheatmasstransfer.2018.06.073
31. Li W, Chen X, Yan L, et al (2018) Additive manufacturing of a new Fe-Cr-Ni alloy with gradually changing compositions with elemental powder mixes and thermodynamic calculation. Int J Adv Manuf Technol 95:1013–1023. https://doi.org/10.1007/s00170-017-1302-1
32. Zhang Y, Liu H, Mo J, et al (2018) Atomic-scale structural evolution in selective laser melting of Cu50Zr50 metallic glass. Comput Mater Sci 150:62–69. https://doi.org/10.1016/j.commatsci.2018.03.072
33. Sorkin A, Tan JL, Wong CH (2017) Multi-material modelling for selective laser melting. Procedia Eng 216:51–57. https://doi.org/10.1016/j.proeng.2018.02.088

34. Tang C, Li B, Wong CH (2014) Selective laser melting of metal powders studied by molecular dynamics simulation. Proc Int Conf Prog Addit Manuf 0:291–296. https://doi.org/10.3850/978-981-09-0446-3_014
35. Zhang Y, Liu H, Mo J, et al (2019) Atomic-level crystallization in selective laser melting fabricated Zr-based metallic glasses. Phys Chem Chem Phys 21:12406–12413. https://doi.org/10.1039/c9cp02181g
36. Bonny G, Castin N, Terentyev D (2013) Interatomic potential for studying ageing under irradiation in stainless steels: The FeNiCr model alloy. Model Simul Mater Sci Eng 21:. https://doi.org/10.1088/0965-0393/21/8/085004
37. Zhang X, Li B, Liu HX, et al (2019) Atomic simulation of melting and surface segregation of ternary Fe-Ni-Cr nanoparticles. Appl Surf Sci 465:871–879. https://doi.org/10.1016/j.apsusc.2018.09.257
38. Pal S, Meraj M (2016) Structural evaluation and deformation features of interface of joint between nano-crystalline Fe-Ni-Cr alloy and nano-crystalline Ni during creep process. Mater Des 108:168–182. https://doi.org/10.1016/j.matdes.2016.06.086
39. Sak-Saracino E, Urbassek HM (2016) The α → γ transformation of an $Fe_{1-x}Cr_x$ alloy: A molecular-dynamics approach. Int J Mod Phys C 27:1–7. https://doi.org/10.1142/S0129183116501242
40. Wu C, Lee BJ, Su X (2017) Modified embedded-atom interatomic potential for Fe-Ni, Cr-Ni and Fe-Cr-Ni systems. Calphad Comput Coupling Phase Diagr Thermochem 57:98–106. https://doi.org/10.1016/j.calphad.2017.03.007
41. Bahramyan M, Mousavian RT, Brabazon D (2020) Study of the plastic deformation mechanism of TRIP-TWIP high entropy alloys at the atomic level. Int J Plast 127:102649. https://doi.org/10.1016/j.ijplas.2019.102649
42. Afkham Y, Bahramyan M, Mousavian RT, Brabazon D (2017) Tensile properties of AlCrCoFeCuNi glassy alloys: A molecular dynamics simulation study. Mater Sci Eng A 698:143–151. https://doi.org/10.1016/j.msea.2017.05.057
43. Sharma A, Singh P, Johnson DD, et al (2016) Atomistic clustering-ordering and high-strain deformation of an $Al_{0.1}$CrCoFeNi high-entropy alloy. Sci Rep 6:1–11. https://doi.org/10.1038/srep31028
44. Sharma A, Balasubramanian G (2017) Dislocation dynamics in $Al_{0.1}$CoCrFeNi high-entropy alloy under tensile loading. Intermetallics 91:31–34. https://doi.org/10.1016/j.intermet.2017.08.004
45. Sharma A, Deshmukh SA, Liaw PK, Balasubramanian G (2017) Crystallization kinetics in $Al_xCrCoFeNi$ (0 ≤ x ≤ 40) high-entropy alloys. Scr Mater 141:54–57. https://doi.org/10.1016/j.scriptamat.2017.07.024
46. Sharma A, Singh R, Liaw PK, Balasubramanian G (2017) Cuckoo searching optimal composition of multicomponent alloys by molecular simulations. Scr Mater 130:292–296. https://doi.org/10.1016/j.scriptamat.2016.12.022
47. Aitken ZH, Sorkin V, Zhang YW (2019) Atomistic modeling of nanoscale plasticity in high-entropy alloys. J Mater Res 34:1509–1532. https://doi.org/10.1557/jmr.2019.50
48. Jafary-Zadeh M, Aitken ZH, Tavakoli R, Zhang YW (2018) On the controllability of phase formation in rapid solidification of high entropy alloys. J Alloys Compd 748:679–686. https://doi.org/10.1016/j.jallcom.2018.03.165
49. 3D Metal Printing / Additive Manufacturing : Fonon Corporation. https://www.fonon.us/3d-metal-printing/. Accessed 3 Feb 2022
50. Plimpton S (1997) Short-Range Molecular Dynamics. J Comput Phys 117:1–42
51. Stukowski A (2010) Visualization and analysis of atomistic simulation data with OVITO-the Open Visualization Tool. Model Simul Mater Sci Eng 18:. https://doi.org/10.1088/0965-0393/18/1/015012

52. Wang HP, Zheng CH, Zou PF, et al (2018) Density determination and simulation of Inconel 718 alloy at normal and metastable liquid states. J Mater Sci Technol 34:436–439. https://doi.org/10.1016/j.jmst.2017.10.014
53. Kiss AM, Fong AY, Calta NP, et al (2019) Laser-Induced Keyhole Defect Dynamics during Metal Additive Manufacturing. Adv Eng Mater 21:. https://doi.org/10.1002/adem.201900455
54. Li L, Peng G, Wang J, et al (2019) Numerical and experimental study on keyhole and melt flow dynamics during laser welding of aluminium alloys under subatmospheric pressures. Int J Heat Mass Transf 133:812–826. https://doi.org/10.1016/j.ijheatmasstransfer.2018.12.165
55. Bayat M, Thanki A, Mohanty S, et al (2019) Keyhole-induced porosities in Laser-based Powder Bed Fusion (L-PBF) of Ti6Al4V: High-fidelity modelling and experimental validation. Addit Manuf 30:100835. https://doi.org/10.1016/j.addma.2019.100835
56. Wang H, Zou Y (2019) Microscale interaction between laser and metal powder in powder-bed additive manufacturing: Conduction mode versus keyhole mode. Int J Heat Mass Transf 142:. https://doi.org/10.1016/j.ijheatmasstransfer.2019.118473
57. Zeng Z, Zhao J, Zhou X, et al (2019) Thermal stability of Al-Cu-Fe-Cr-Ni high entropy alloy bulk and nanoparticle structure: A molecular dynamics perspective. Chem Phys 517:126–130. https://doi.org/10.1016/j.chemphys.2018.10.009
58. XIAO F, xiao LIU L, hui YANG R, et al (2008) Surface tension of molten Ni-(Cr, Co, W) alloys and segregation of elements. Trans Nonferrous Met Soc China Engl Ed 18:1184–1188. https://doi.org/10.1016/S1003-6326(08)60202-2
59. Zhu Y, Xiang Y, Schulz K (2016) The role of dislocation pile-up in flow stress determination and strain hardening. Scr Mater 116:53–56. https://doi.org/10.1016/j.scriptamat.2016.01.025

| | List of Figure Captions |
|---|---|
| **Figure 1** | MD simulation model (a) for layer resolution 6. Top View at (b) beginning and (c) end of LPBF process for layer resolution 1. The yellow circle represents the laser, while the white box indicates the region to be cut off from the rest of the simulation cell to perform the uniaxial tensile test. |
| **Figure 2** | Illustration of the cut-off section from layer resolution 1 simulation cell and application of non-periodic uniaxial tensile test along the Z direction. |
| **Figure 3** | (a) Control simulation of strain-controlled uniaxial tensile test for an equimolar FeNiCr MEA block that hasn't undergone any heat treatment. Loading and unloading the block reveals that it is within the elastic limit up to 8.28 GPa at 15.6% strain. (b) Strained blocks show that necking starts to occur beyond this limit. |
| **Figure 4** | Illustration of layer addition mechanism for various layer resolutions. For layer resolution 1 and resolution 3, new rows of powder are added after cooling the previous layer's melt. In the case of layer resolution 6, all 6 rows of powders are melted at the same time using the laser. |
| **Figure 5** | Common neighbor analysis (CNA) of sectional view and 3D view of element distribution of the post-cooled structures. |
| **Figure 6** | (a) Stress-strain relationship of cut-off blocks obtained from uniaxial tensile tests, (b) changes in simulation process time and shrinkage, (c) variation in properties of block for 100 µW laser power, 300 K bed temperature, and 0.5 Å/ps scan speed with increasing layer resolution. |
| **Figure 7** | Sectional view of the common neighbor analysis (CNA) for various laser powers portraying the formation of laser-induced defects. |
| **Figure 8** | (a) Stress-strain relationship of cut-off blocks obtained from uniaxial tensile tests, (b) changes in simulation process time and shrinkage, (c) variation in properties of block for layer resolution 6, 300 K bed temperature, and 0.5 Å/ps scan speed with increasing laser power. |
| **Figure 9** | (a) Stress-strain relationship of cut-off blocks obtained from uniaxial tensile tests, (b) changes in simulation process time and shrinkage, (c) variation in properties of block for layer resolution 6, 100 µW laser power, and 0.5 Å/ps scan speed with increasing bed temperature. |
| **Figure 10** | Sectional view of the post-cooling structures procured for different bed temperatures. |
| **Figure 11** | (a) Stress-strain relationship of cut-off blocks obtained from uniaxial tensile tests, (b) changes in simulation process time and shrinkage, (c) variation in properties of block for layer resolution 6, 100 µW laser power, and 300 K bed temperature with decreasing scanning speed. |
| **Figure 12** | Element distribution and dislocation analysis (DXA) of the blocks obtained for different scan speeds. For clarity, only the atoms in HCP stacking faults and dislocation lines are shown in DXA. |
| **Figure 13** | Bar graph showing the percentage increase ultimate tensile strength (UTS) and process time of the best strength conditions for each process parameter, as compared to that of the baseline simulation C. |

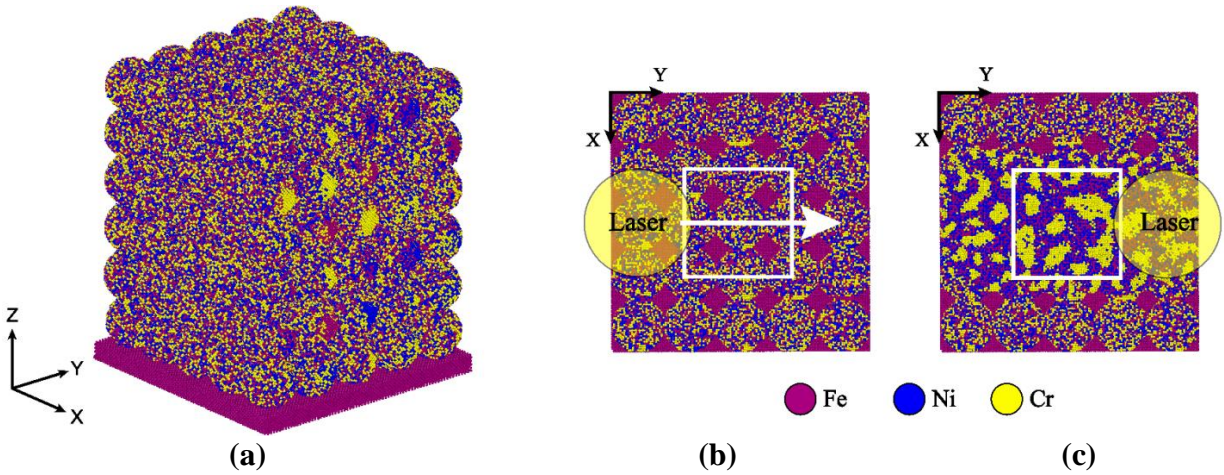

**Figure 1.** MD simulation model (a) for layer resolution 6. Top View at (b) beginning and (c) end of LPBF process for layer resolution 1. The yellow circle represents the laser, while the white box indicates the region to be cut off from the rest of the simulation cell to perform the uniaxial tensile test.

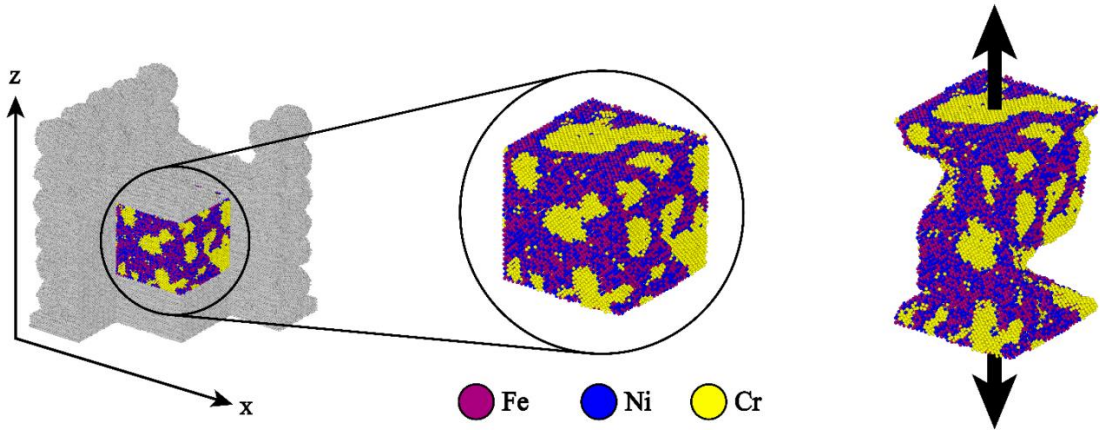

**Figure 2.** Illustration of the cut-off section from layer resolution 1 simulation cell and application of non-periodic uniaxial tensile test along the Z direction.

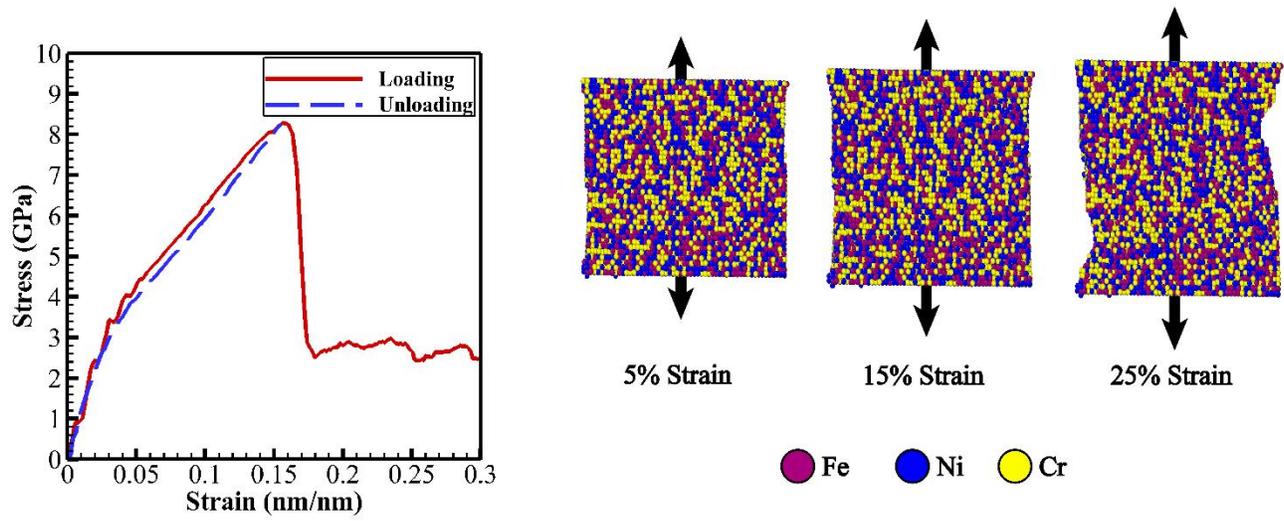

**Figure 3.** Control simulation of strain-controlled uniaxial tensile test for an equimolar FeNiCr MEA block that hasn't undergone any heat treatment. Loading and unloading the block reveals that it is within the elastic limit up to 8.28 GPa at 15.6% strain. Strained blocks show that necking starts to occur beyond this limit.

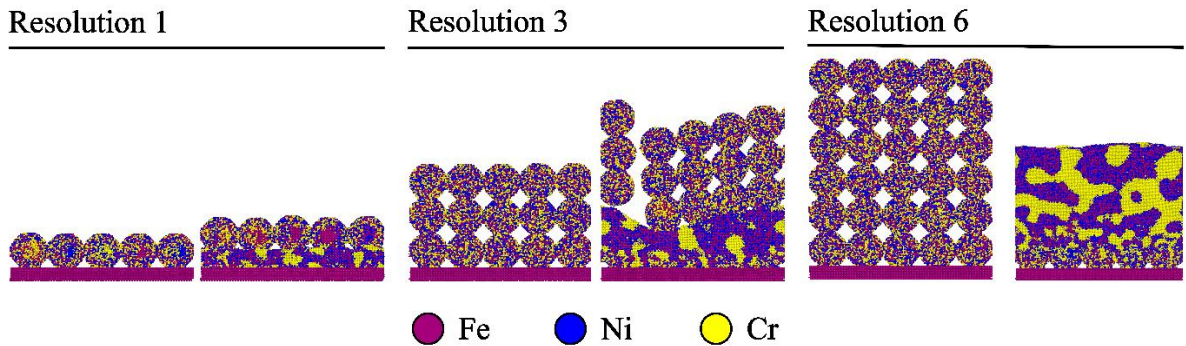

**Figure 4.** Illustration of layer addition mechanism for various layer resolutions. For layer resolution 1 and resolution 3, new rows of powder are added after cooling the previous layer's melt. In the case of layer resolution 6, all 6 rows of powders are melted at the same time using the laser.

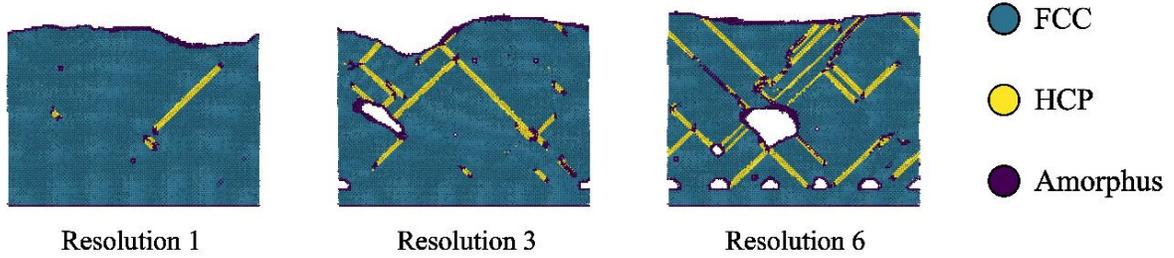

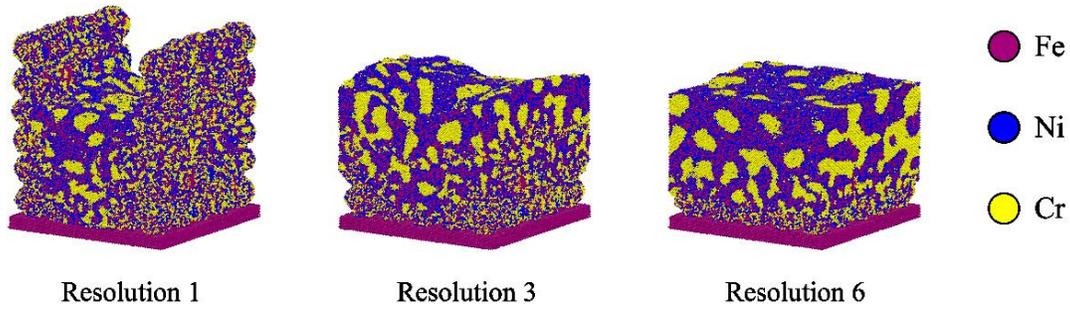

**Figure 5.** Common neighbor analysis (CNA) of sectional view and 3D view of element distribution of the post-cooled structures.

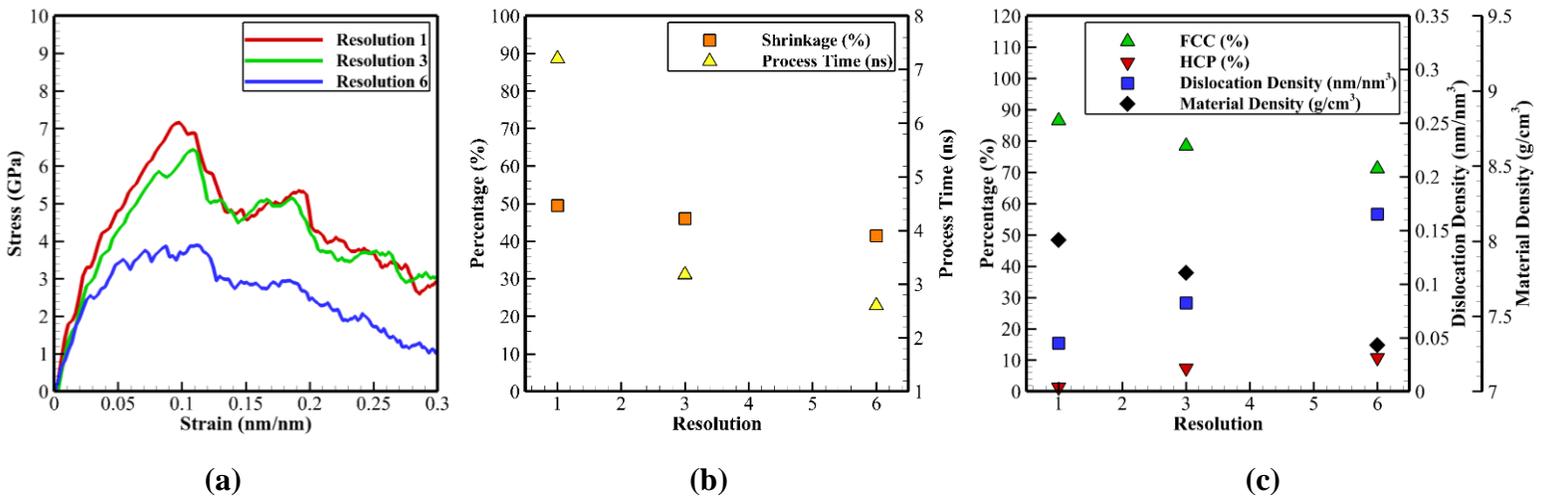

**Figure 6.** (a) Stress-strain relationship of cut-off blocks obtained from uniaxial tensile tests, (b) changes in simulation process time and shrinkage, (c) variation in properties of block for 100 µW laser power, 300 K bed temperature, and 0.5 Å/ps scan speed with increasing layer resolution.

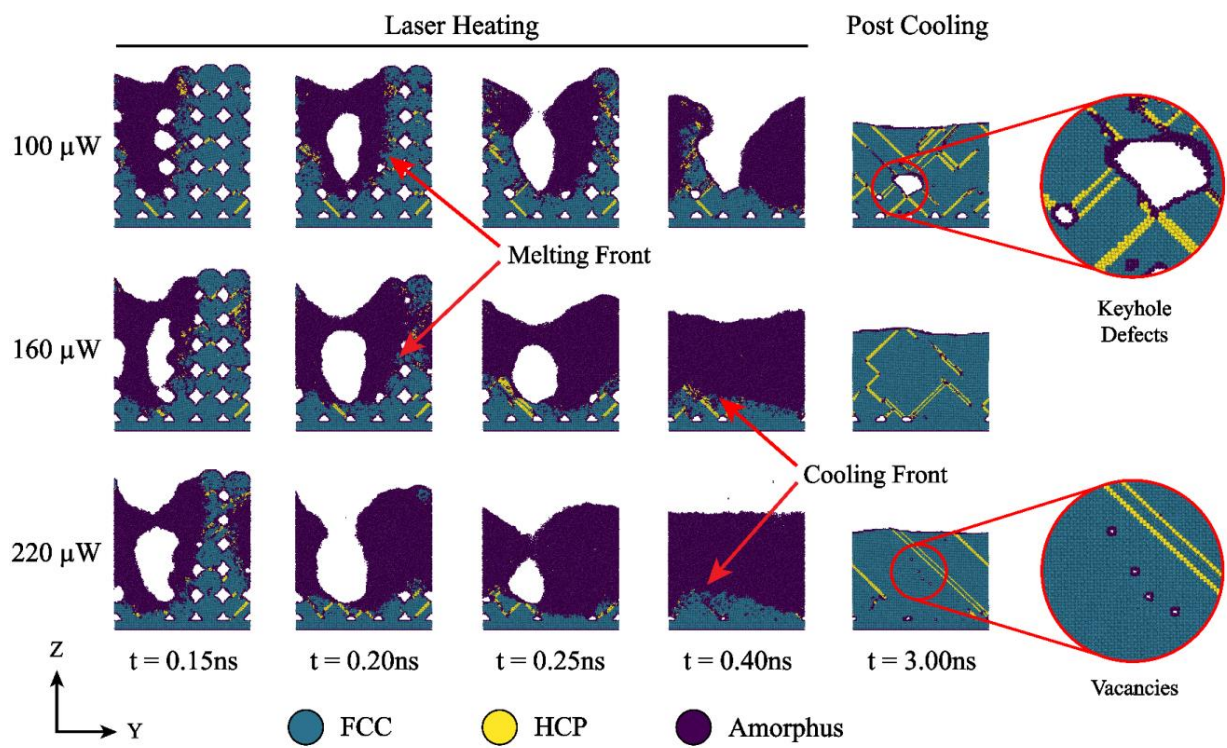

**Figure 7.** Sectional view of the common neighbor analysis (CNA) for various laser powers portraying the formation of laser-induced defects.

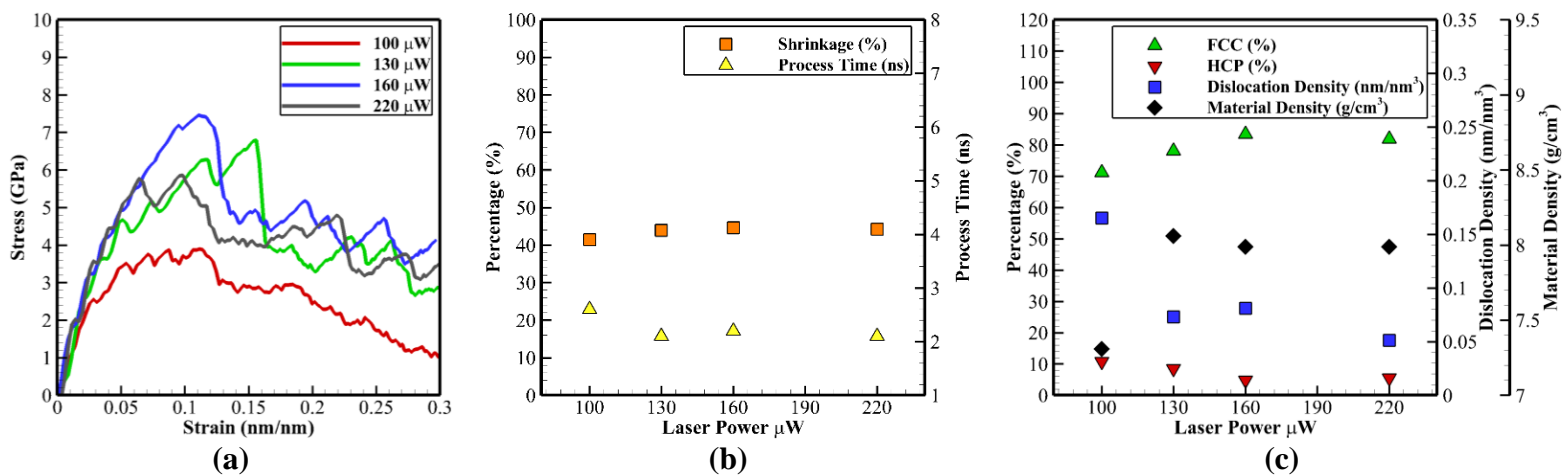

**Figure 8.** (a) Stress-strain relationship of cut-off blocks obtained from uniaxial tensile tests, (b) changes in simulation process time and shrinkage, (c) variation in properties of block for layer resolution 6, 300 K bed temperature, and 0.5 Å/ps scan speed with increasing laser power.

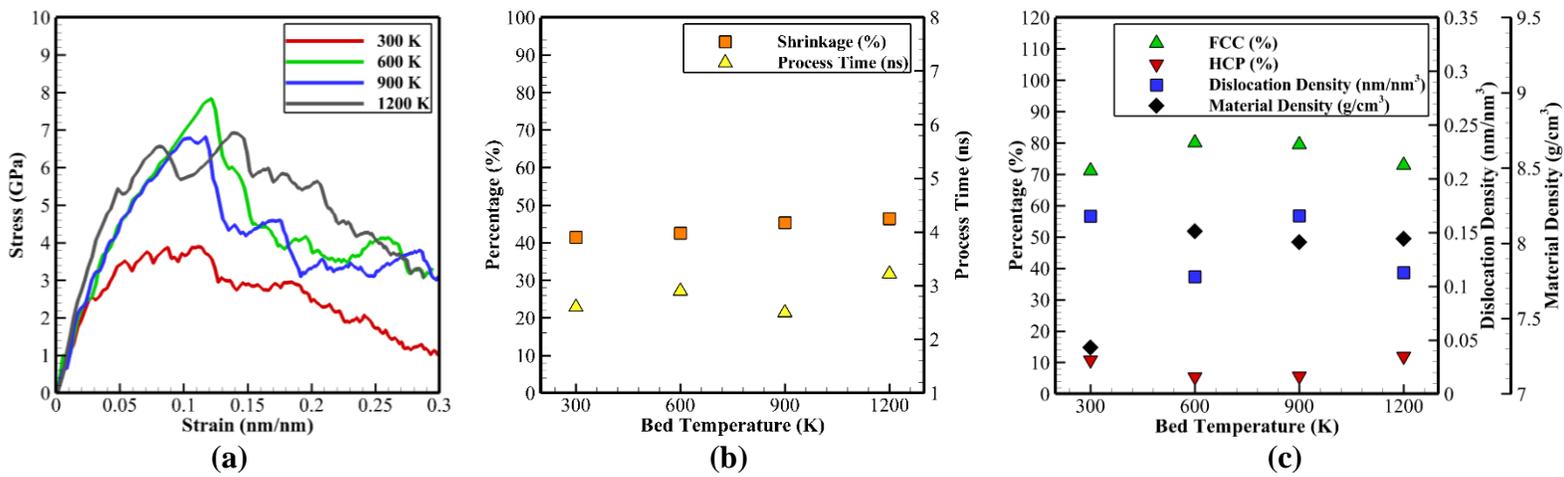

**Figure 9.** (a) Stress-strain relationship of cut-off blocks obtained from uniaxial tensile tests, (b) changes in simulation process time and shrinkage, (c) variation in properties of block for layer resolution 6, 100 µW laser power, and 0.5 Å/ps scan speed with increasing bed temperature.

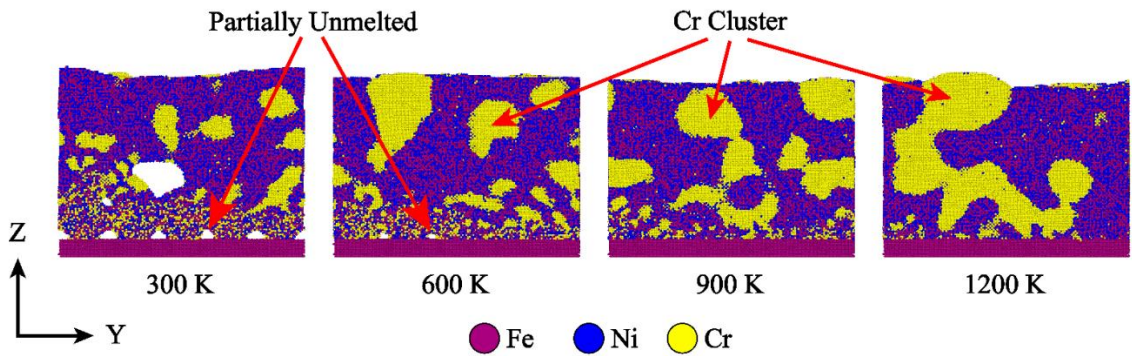

**Figure 10.** Sectional view of the post-cooling structures procured for different bed temperatures.

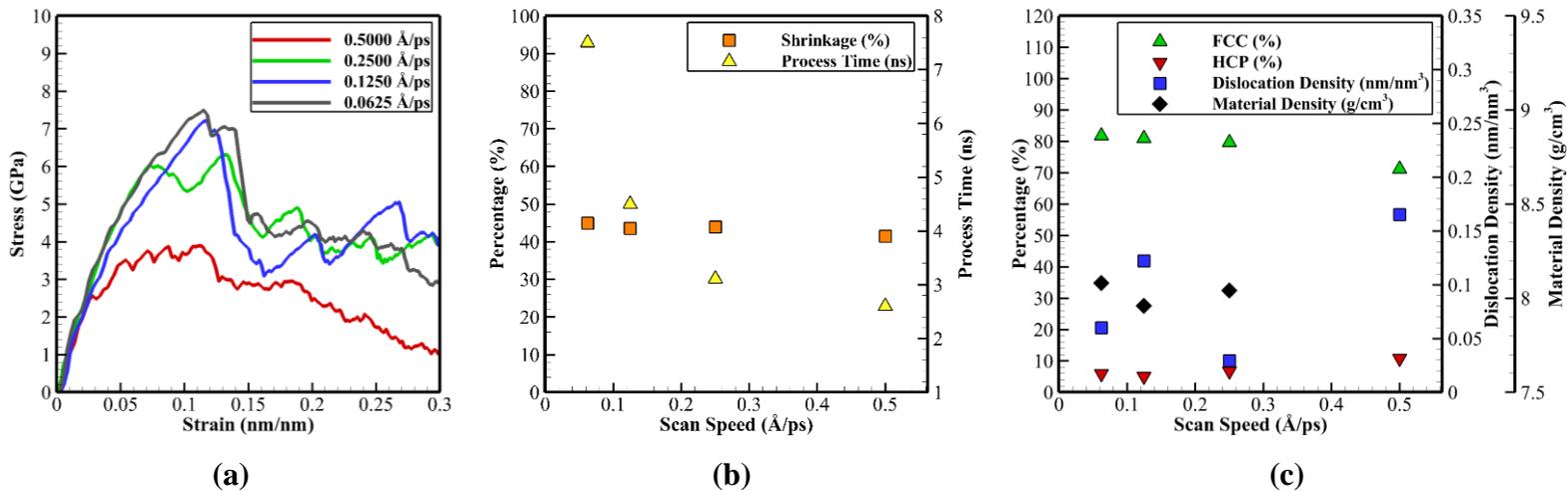

**(a)**                                    **(b)**                                   **(c)**

**Figure 11.** (a) Stress-strain relationship of cut-off blocks obtained from uniaxial tensile tests, (b) changes in simulation process time and shrinkage, (c) variation in properties of block for layer resolution 6, 100 µW laser power, and 300 K bed temperature with decreasing scanning speed.

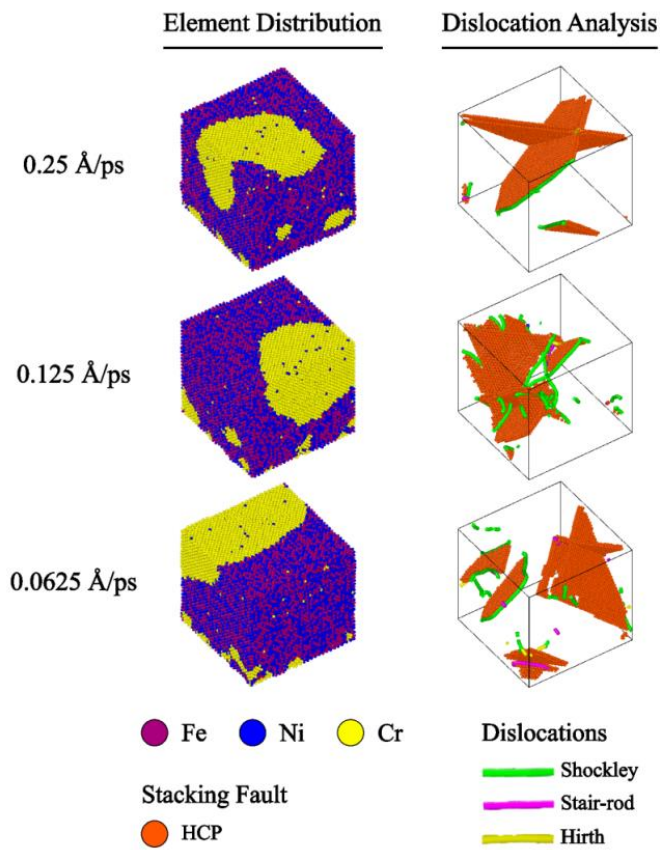

**Figure 12.** Element distribution and dislocation analysis (DXA) of the blocks obtained for different scan speeds. For clarity, only the atoms in HCP stacking faults and dislocation lines are shown in DXA.

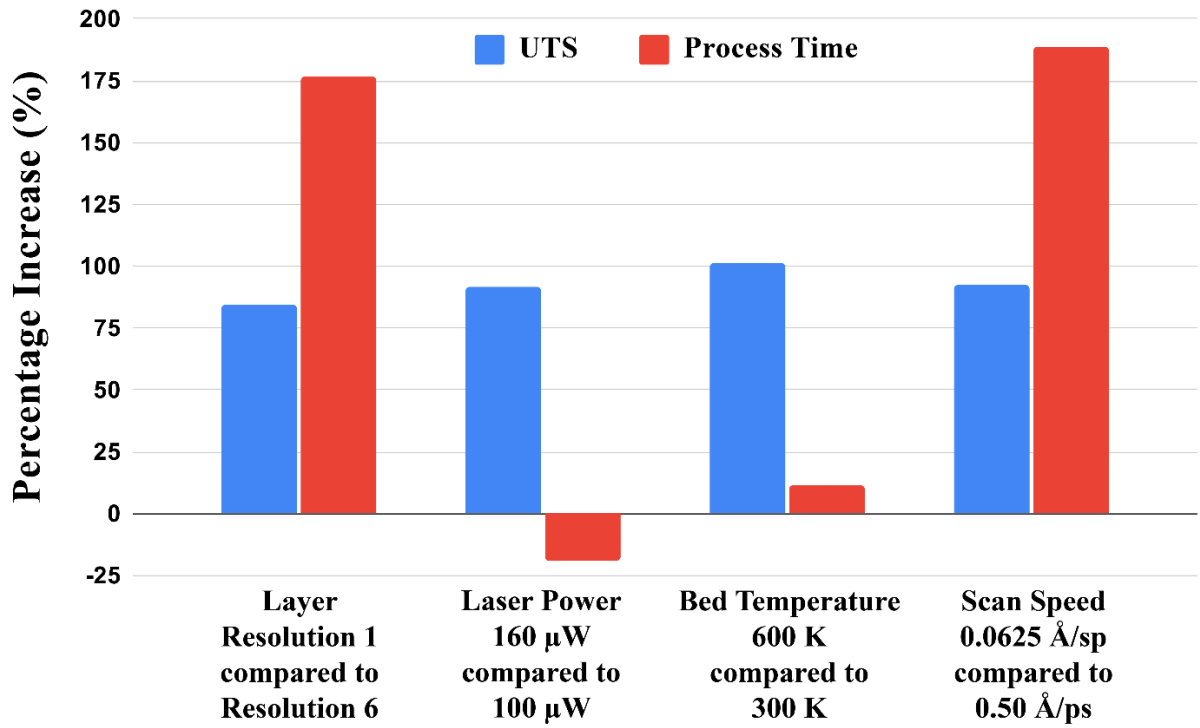

**Figure 13.** Bar graph showing the percentage ultimate tensile strength (UTS) and process time of the best strength conditions for each process parameter, as compared to that of the baseline simulation C.